\documentclass[a4paper,nobibnotes,nofootinbib,superscriptaddress]{revtex4}

\usepackage{amssymb}
\usepackage{amsmath}
\usepackage{epsfig}

\newcommand{\be}{\begin{equation}}
\newcommand{\ee}{\end{equation}}
\newcommand{\bea}{\begin{eqnarray}}
\newcommand{\eea}{\end{eqnarray}}
\newcommand{\nn}{\nonumber}

\newcommand{\g}{\gamma}
\newcommand{\f}{\frac}

\newcommand{\intc}[1]{{\int\frac{d#1}{2i\pi}}}
\newcommand\lr[1]{{\left({#1}\right)}}

\begin{document}

\title{Gaps between jets at hadron colliders in the next-to-leading BFKL framework}
\author{F. Chevallier}\email{florent.chevallier@cea.fr}
\affiliation{IRFU/Service de physique des particules, CEA/Saclay, 91191 Gif-sur-Yvette cedex, France}
\author{O. Kepka}\email{kepkao@fzu.cz}
\affiliation{IRFU/Service de physique des particules, CEA/Saclay, 91191 Gif-sur-Yvette cedex, France}
\affiliation{IPNP, Faculty of Mathematics and Physics, Charles University, Prague, Czech Republic}
\affiliation{Center for Particle Physics, Institute of Physics, Academy of Science, Prague, Czech Republic}
\author{C. Marquet}\email{cyrille@phys.columbia.edu}
\affiliation{Institut de Physique Th{\'e}orique, CEA/Saclay, 91191 Gif-sur-Yvette cedex, France}
\affiliation{Department of Physics, Columbia University, New York, NY 10027, USA}
\author{C. Royon}\email{royon@hep.saclay.cea.fr}
\affiliation{IRFU/Service de physique des particules, CEA/Saclay, 91191 Gif-sur-Yvette cedex, France}

\begin{abstract}

We investigate diffractive events in hadron-hadron collisions, in which two jets are produced and separated by a large
rapidity gap. In perturbative QCD, the hard color-singlet object exchanged in the $t-$channel, and responsible for the
rapidity gap, is the Balitsky-Fadin-Kuraev-Lipatov (BFKL) Pomeron. We perform a phenomenological study including
the corrections due to next-to-leading logarithms (NLL). Using a renormalisation-group improved NLL kernel, we show that the BFKL predictions are in good agreement with the Tevatron data, and present predictions which could be tested at the LHC.

\end{abstract}

\maketitle

\section{Introduction}

The discovery of diffractive interactions in hard processes at hadron colliders has triggered a large effort
devoted to understand rapidity gaps in processes with a hard scale. Since events with a gap between two jets
were observed in $p\bar{p}$ collisions at the Tevatron about 10 years ago \cite{d0jgj}, many theoretical
works have investigated these processes in QCD. While describing diffractive processes in QCD has been a
challenge for many years, one should be able to understand hard diffractive events with perturbative methods.

In a hadron-hadron collision, a jet-gap-jet event features a large rapidity gap with a high$-E_T$ jet on each side
($E_T\!\gg\!\Lambda_{QCD}$). Across the gap, the object exchanged in the $t\!-\!channel$ is color singlet and carries a large momentum transfer, and when the rapidity gap is sufficiently large the natural candidate in perturbative QCD is the
Balitsky-Fadin-Kuraev-Lipatov (BFKL) Pomeron \cite{bfkl}. Of course the total energy of the collision $\sqrt{s}$
should be big ($\sqrt{s}\gg E_T$) in order to get jets and a large rapidity gap.

To compute the jet-gap-jet process in the BFKL framework, the problem of coupling the BFKL Pomeron to partons,
instead of colorless particles in the standard approach, has to be solved first. Indeed, one usually uses the fact
that impact factors vanish when hooked to gluons with no transverse momentum, a property of colorless impact factors.
For instance, this is what allows to turn the Feynman diagram calculation into a conformal invariant Green function \cite{lipatov}.
However, with colored particules, this BFKL Green function cannot be used and should be modified accordingly.
The Mueller-Tang (MT) prescription \cite{muellertang} is widely used in the literature.

On the phenomenological side, the original MT calculation is not sufficient to describe the data. A first try to improve it
was attempted \cite{cfl} by also taking into account the soft interactions which in hadron-hadron collisions can destroy
the rapidity gap. An agreement was only obtained if the leading-logarithmic (LL) BFKL calculation was done with a fixed
value of the coupling constant $\alpha_s,$ which is not satisfactory. Moreover, it is also known that next-to-leading
logarithmic (NLL) BFKL corrections can be large. In \cite{rikard}, it was shown that a good description of the data could
be obtained when some NLL corrections were numerically taken into account in an effective way \cite{fakenll}, but the full
NLL-BFKL kernel \cite{nllbfkl} could still not be implemented. As a result, these tests on the relevance of the BFKL
dynamics were not conclusive.

On the theoretical side, it is known that NLL corrections to the LL-BFKL predictions can be large due to the appearance
of spurious singularities in contradiction with renormalization-group requirements. However it has been realised \cite{salam,CCS}
that a renormalisation-group improved NLL regularisation can solve the singularity problem and lead to reasonable
NLL-BFKL kernels (see also \cite{singnll} for different approaches). This motivates our phenomenological study of
jet-gap-jet events in the NLL-BFKL framework. One goal is to motivate further measurements at the Tevatron and the LHC.

Such NLL-BFKL phenomenological investigations have been devoted to the proton structure function \cite{nllf2},
forward jet production in deep inelastic scattering \cite{nllfjus,nllfjthem}, and Mueller-Navelet jets in hadron-hadron
collisions \cite{nllmnjus,nllmnjthem}. While for the structure function analysis the NLL corrections did not really improve the
BFKL description, it was definitively the case in the forward jet analysis, and they play a non-negligible role for the
Mueller-Navelet jet predictions. Therefore, it is natural to investigate what happens for jet-gap-jet events, and what is
the magnitude of the NLL corrections with respect to the LL-BFKL results for this observable.

The plan of the paper is the following. In section II, we introduce the phenomenological NLL-BFKL formulation of the
jet-gap-jet cross section. In section III, we compare the LL- and NLL-BFKL calculations, and confront them with the
Tevatron measurements. In section V, we present predictions for the jet-gap-jet cross section at the LHC. Section V
is devoted to conclusions and outlook.

\section{The jet-gap-jet cross section in the BFKL framework}

In a hadron-hadron collision, the production of two jets with a gap in rapidity between them is represented in Fig.1,
with the different kinematic variables. We denote $\sqrt{s}$ the total energy of the collision,
$p_T$ and $-p_T$ the transverse momenta of the two jets and $x_1$ and $x_2$ their longitudinal
fraction of momentum with respect to the incident hadrons as indicated on the figure. The rapidity gap
between the two jets is $\Delta\eta\!=\!\ln(x_1x_2s/p_T^2).$ We write the cross section in the following form
\begin{equation}
\frac{d \sigma^{pp\to XJJY}}{dx_1 dx_2 dp_T^2} = {\cal S}f_{eff}(x_1,p_T^2)f_{eff}(x_2,p_T^2)
\frac{d \sigma^{gg\rightarrow gg}}{dp_T^2},
\label{jgj}\end{equation}
where the functions $f_{eff}(x,p_T^2)$ are effective parton distributions that resum the leading logarithms $\log(p_T^2/\Lambda_{QCD}^2).$ They have the form
\be
f_{eff}(x,\mu^2)=g(x,\mu^2)+\f{C_F^2}{N_c^2}\lr{q(x,\mu^2)+\bar{q}(x,\mu^2)}\ ,
\label{pdfs}
\ee
where $g$ (respectively $q$, $\bar{q}$) is the gluon (respectively quark, antiquark) distribution function in the incoming hadrons. Even though
the process we consider involves moderate values of $x_1$ and $x_2$ and the perturbative scale $p_T^2\gg\Lambda_{QCD}^2,$ which we 
have chosen as the factorization scale, the cross section \eqref{jgj} does not obey collinear factorization. This is due to 
possible secondary soft interactions between the colliding hadrons which can fill the rapidity gap. Therefore, in \eqref{jgj}, 
the collinear factorization of the parton distributions $f_{eff}$ is corrected with the so-called gap survival probability ${\cal S}:$ 
since the soft interactions happen on much longer time scales, they are factorized from the hard part $d \sigma^{gg\rightarrow gg}/dp_T^2.$ 
This cross section is given by
\begin{equation}
\frac{d \sigma^{gg\rightarrow gg}}{dp_T^2}=\frac{1}{16\pi}\left|A(\Delta\eta,p_T^2)\right|^2
\end{equation}
in terms of the $gg\to gg$ scattering amplitude $A(\Delta\eta,p_T^2).$ The two measured jets are initiated by the final-state
gluons and we are neglecting possible hadronization effects.

In the following, we consider the high energy limit in which the rapidity gap $\Delta\eta$ is assumed to be very large.
The BFKL framework allows to compute the $gg\to gg$ amplitude in this regime, and the result is known up to NLL accuracy.
Since in this calculation the BFKL Pomeron is coupled to quarks or gluons, the BFKL Green function cannot be used as it is
and should be modified. The transformation proposed in \cite{muellertang} is based on the fact that one should recover
the analiticity of the Feynman diagrams. It was later argued that their prescripion corresponds to a deformed representation
of the BFKL kernel that indeed could be coupled to colored particules and for which the bootstrap relation is fullfiled \cite{barlip}.
Applying the MT prescription at NLL leads to
\begin{equation}
A(\Delta\eta,p_T^2)=\frac{16N_c\pi\alpha_s^2}{C_Fp_T^2}\sum_{p=-\infty}^\infty\intc{\g}
\frac{[p^2-(\g-1/2)^2]\exp\left\{\bar\alpha(p_T^2)\chi_{eff}[2p,\g,\bar\alpha(p_T^2)] \Delta \eta\right\}}
{[(\g-1/2)^2-(p-1/2)^2][(\g-1/2)^2-(p+1/2)^2]} 
\label{jgjnll}\end{equation}
with the complex integral running along the imaginary axis from $1/2\!-\!i\infty$ 
to $1/2\!+\!i\infty,$ and with only even conformal spins contributing to the sum \cite{leszek}.
The running coupling is given by
\be
\bar\alpha(p_T^2)=\alpha_s(p_T^2)N_c/\pi=
\left[b\log\lr{p_T^2/\Lambda_{QCD}^2}\right]^{-1}\ ,
\hspace{1cm}b=\f{11N_c-2N_f}{12N_c}\ .\label{runc}\ee

\begin{figure}[t]
\begin{center}
\epsfig{file=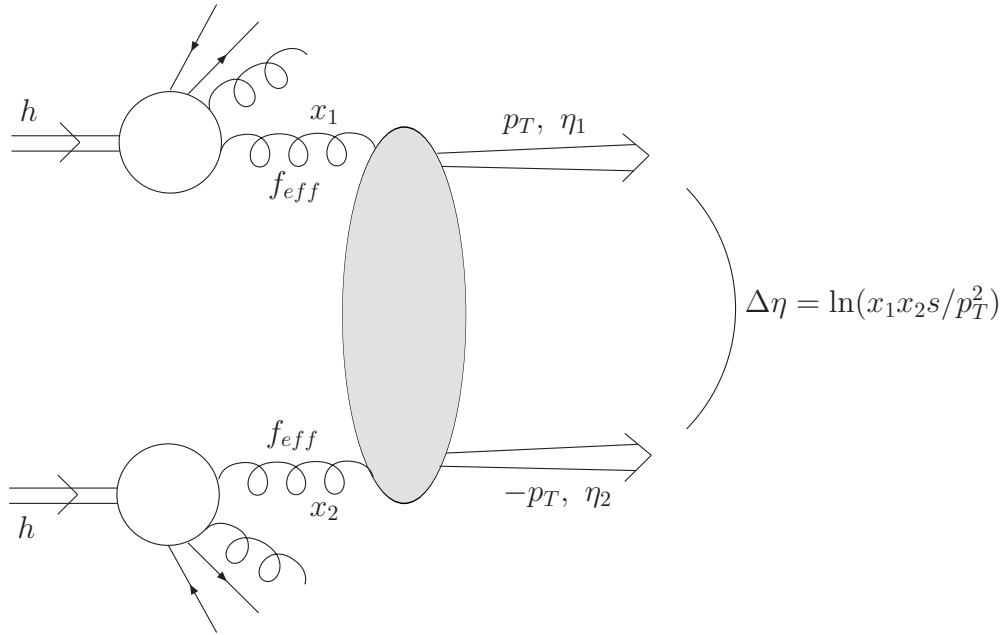,width=13cm}
\caption{Production of two jets separated by a large rapidity gap in a hadron-hadron collision. The kinematic variables of the problem are displayed. $s$ is the total energy squared of the collision, $p_T$ ($\eta_1$) and $-p_T$ ($\eta_2$) are the transverse momenta (rapidities) of the jets and $x_1$ and $x_2$ are their longitudinal momentum fraction with respect to the incident hadrons. $\Delta\eta$ is the size of rapidity gap between the jets.}
\end{center}
\end{figure}

Let us give some more details on formula \eqref{jgjnll}. The NLL-BFKL effects are phenomenologically taken into account by the
effective kernels $\chi_{eff}(p,\g,\bar\alpha).$ For $p=0,$ the scheme-dependent
NLL-BFKL kernels provided by the regularisation procedure $\chi_{NLL}\lr{\g,\omega}$ depend
on $\omega,$ the Mellin variable conjugate to $\exp(\Delta\eta).$ In each case, the NLL kernels obey a {\it consistency condition} \cite{salam} which allows to reformulate the problem in terms of $\chi_{eff}(\g,\bar\alpha).$ The effective kernel
$\chi_{eff}(\g,\bar\alpha)$ is obtained from the NLL kernel $\chi_{NLL}\lr{\g,\omega}$ by solving the implicit equation
$\chi_{eff}=\chi_{NLL}\lr{\g,\bar\alpha\ \chi_{eff}}$ as a solution of the consistency condition.

In \cite{nllmnjus,nllmnjthem}, the regularisation procedure has been extended to non-zero conformal spins and the kernel $\chi_{NLL}\lr{p,\g,\omega}$ was obtained,
the formulae needed to compute it can be found in the appendix of \cite{nllmnjus} (in the present study we shall use the S4 scheme in which $\chi_{NLL}$ is supplemented by an explicit $\bar\alpha$ dependence, the results in the case of the S3 scheme are similar).
Then the effective
kernels $\chi_{eff}(p,\g,\bar\alpha)$ are obtained from the NLL kernel by solving the implicit equation:
\be
\chi_{eff}=\chi_{NLL}\lr{p,\g,\bar\alpha\ \chi_{eff}}\ .
\label{eff}
\ee
It is important to note that in formula \eqref{jgjnll}, we used the leading-order non-forward quark and gluon impact factors.
We point out that the next-to-leading-order impact factors are known \cite{ifnlo}, and that in principle a full NLL analysis is
feasible, but this goes beyond the scope of our study.

By comparison, the LL-BFKL formula is formally the same as \eqref{jgjnll} with the substitutions
\be
\chi_{eff}(p,\g,\bar\alpha)\rightarrow\chi_{LL}(p,\g)
=2\psi(1)-\psi\lr{1-\g+\f{|p|}2}-\psi\lr{\g+\f{|p|}2}\ ,
\hspace{1cm}\bar\alpha(k^2)\rightarrow\bar\alpha=\mbox{const. parameter} ,
\label{chill}\ee
where $\psi(\g)\!=\!d\log\Gamma(\g)/d\g$ is the logarithmic derivative of the Gamma function. In the NLL-BFKL formula,
the value of $\bar\alpha$ is imposed by the scale $p_T^2$ and the variation of $\alpha_S$ is given in Formula (5). 
We shall later test the sensitivity of our results when using $\lambda p_T^2$ and varying 
$\lambda$ between 0.5 and 2. This is done using formula (4) with the appropriate substitutions \cite{nllfjus,nllfjthem}
\bea
\bar\alpha(p_T^2)&\rightarrow&\bar\alpha(\lambda p_T^2)\!+\!b\ \bar\alpha^2(p_T^2)\ln(\lambda)\ ,\nn\\
\Delta\eta&\rightarrow&\Delta\eta-\ln(\lambda)\ .
\eea
By contrast, in the LL-BFKL case that we consider for
comparisons, $\bar\alpha$ is a priori a parameter. We choose to fix it to the value 0.16 obtained in \cite{nllfjus} by
fitting the forward jet data from HERA. This unphysically small value of the coupling (0.16) is indicative of the slower energy
dependence of the forward jet data compared to the LL-BFKL cross section. And in fact, the value $\bar\alpha=0.16$
mimics the slower energy dependence of NLL-BFKL cross section, which in the forward jet case is consistent with the data.
Therefore in both cases one deals with one-parameter formulae: the absolute normalization is not under control. In the NLL case, this is due to the fact the we do not use NLO impact factors, although the correction should be of order one.

\section{Comparaison with Tevatron data}
The D0 collaboration has performed a measurement of the jet-gap-jet event rate as a function of the second
leading jet transverse energy $E_T=|p_T|,$ and also as a function of the rapidity difference
$\Delta\eta$ between the two leading jets \cite{d0jgj}. They requested
two jets reconstructed in the D0 calorimeter with $E_T> 15\ \mbox{GeV}$ for the second leading jet and the following cuts
on the jet rapidities: $1.9<|\eta_{1,2}|<4.1$ and $\eta_1 \eta_2 <0.$ The difference in rapidity between both jets $\Delta\eta$ was imposed to be bigger than 4. The D0 collaboration measured the ratio of that production cross section with a gap of at
least 4 units of rapidity between the jets, to the dijet inclusive cross section. The first measurement performed presented the
jet-gap-jet ratio as a function of the second leading jet $E_T,$ and the second and third measurements displayed the ratio as
a function of $\Delta\eta$ for the low$-E_T$ and high$-E_T$ jet samples respectively (low $E_T$ means
$15<E_T<25\ \mbox{GeV}$ and high $E_T$ means $E_T>30\ \mbox{GeV}$ where $E_T$ means the transverse energy of the leading and the second leading
jets).

In the following, instead of using $x_1$ and $x_2$ we will work with the rapidities of the two jets $\eta_1=\ln(x_1\sqrt{s}/E_T)$ and 
$\eta_2=-\ln(x_2\sqrt{s}/E_T),$ and with the average rapidity $y=(\eta_1+\eta_2)/2:$
\begin{equation}
x_1 = \frac{E_T}{\sqrt s} \exp (\Delta \eta / 2 + y)\ ,\hspace{0.5cm}
x_2 = \frac{E_T}{\sqrt s} \exp (\Delta \eta / 2 - y)\ ,\hspace{0.5cm}
\frac{d\sigma^{pp\to XJJY}}{dyd \Delta\eta dE_T^2}=x_1 x_2 \frac{d\sigma^{pp\to XJJY}}{dx_1dx_2dp_T^2}\ .
\end{equation}
For comparison with the different measurements, we integrate the NLL-BFKL cross section over $y$ and $\Delta\eta,$ or $y$ and $E_T,$
taking into account the cuts detailed above.

\subsection{Effect of non-zero conformal spins and NLL/LL comparison}

\begin{figure}[t]
\begin{center}
\epsfig{file=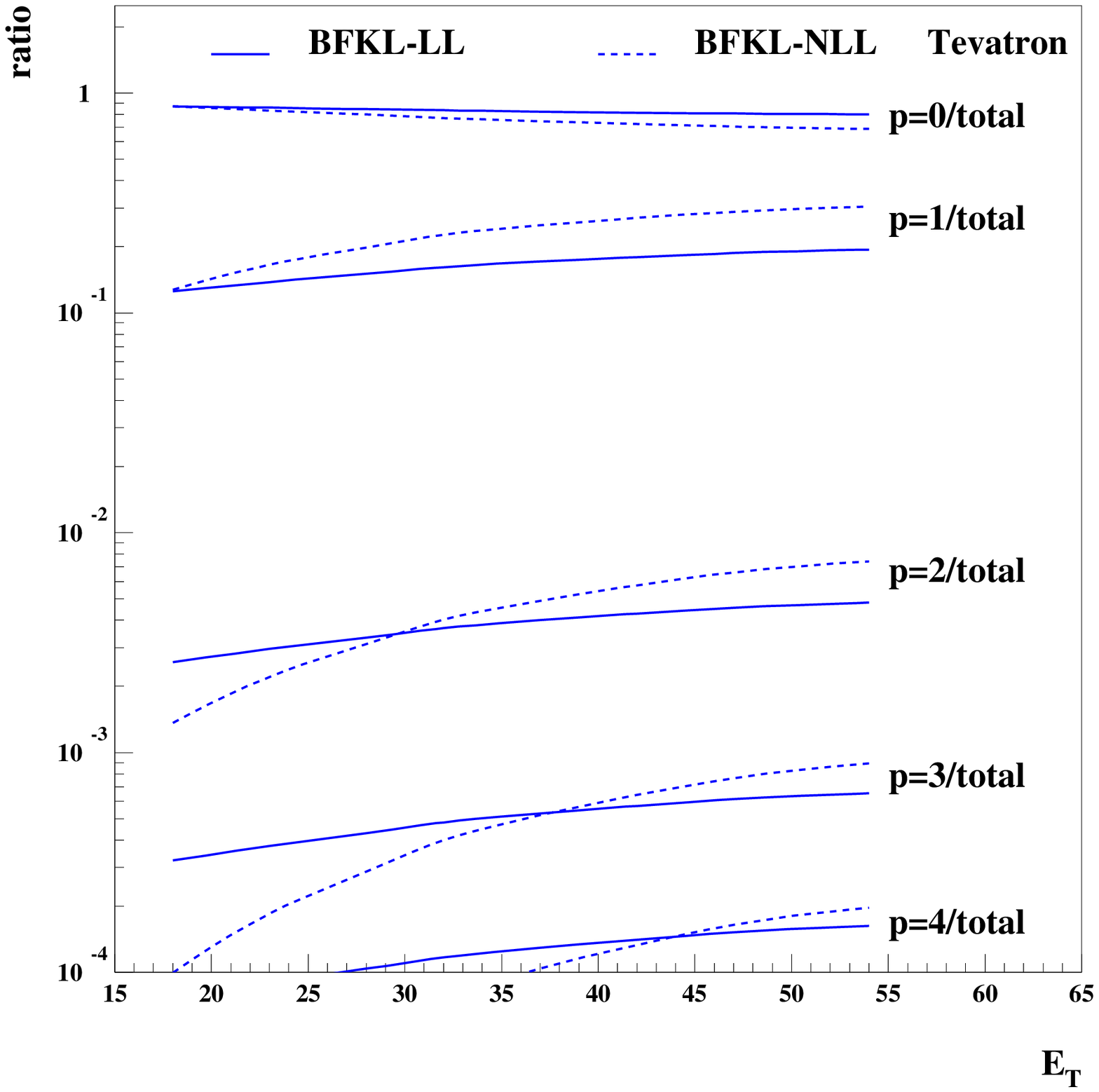,width=5.5cm}
\hfill
\epsfig{file=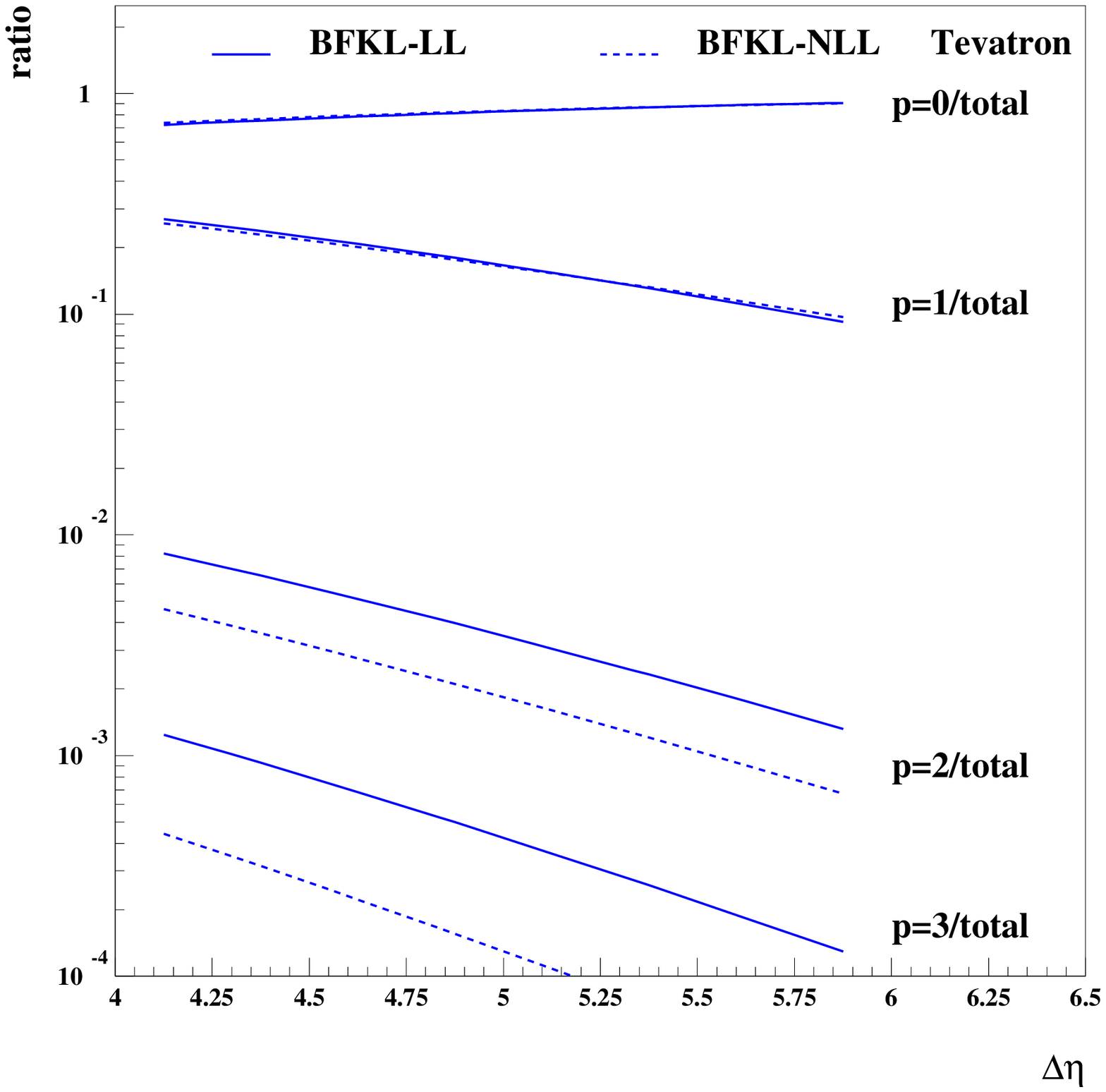,width=5.5cm}
\hfill
\epsfig{file=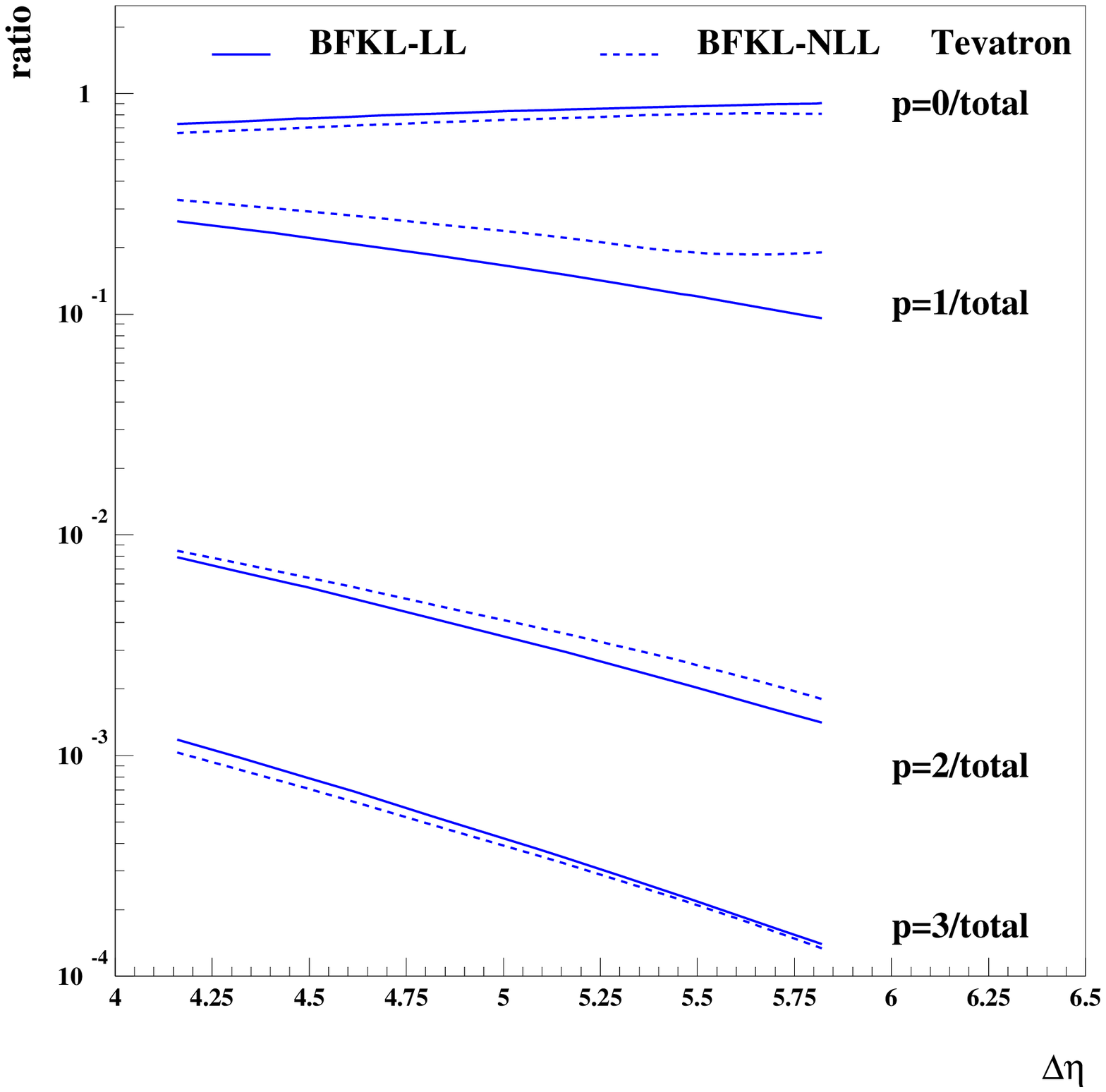,width=5.5cm}
\caption{Ratio between the conformal-spin components $p=0$, 1, 2, 3 and 4, and the total BFKL cross section at LL and NLL accuracy. 
Left plot: as a function of $E_T$ for $\Delta \eta >4;$ center plot: as a function $\Delta\eta$ for the second leading jet with
$15<E_T<25\ \mbox{GeV};$ right plot: as a function $\Delta\eta$ for the second leading jet with $E_T > 30\ \mbox{GeV}.$}
\label{fig2}
\end{center}
\end{figure}

Before comparing our BFKL calculations with the Tevatron measurements, it is worth discussing the effect of taking into account all the 
conformal spins at LL and NLL which was not performed up to now. In Fig.~\ref{fig2}, we give the ratio of the different conformal spin components for 
$p=0$, 1, 2, 3 and 4 with respect to the sum of all components. First we see that the $p=0$ component dominates the cross section, 
however the contribution of the $p=1$ component is large at high $E_T$ and low $\Delta \eta,$ where it can reach up to 30\%. Its
$E_T$ dependence is also not negligible and varies between 10 to 30\% for a tranverse energy between 
15 and 60 GeV. It is also worth to notice that considering only the conformal spin $p=0$ component instead of computing the sum of 
all components leads in general to larger differences at LL than at NLL, especially at high $E_T.$

\begin{figure}[t]
\begin{center}
\epsfig{file=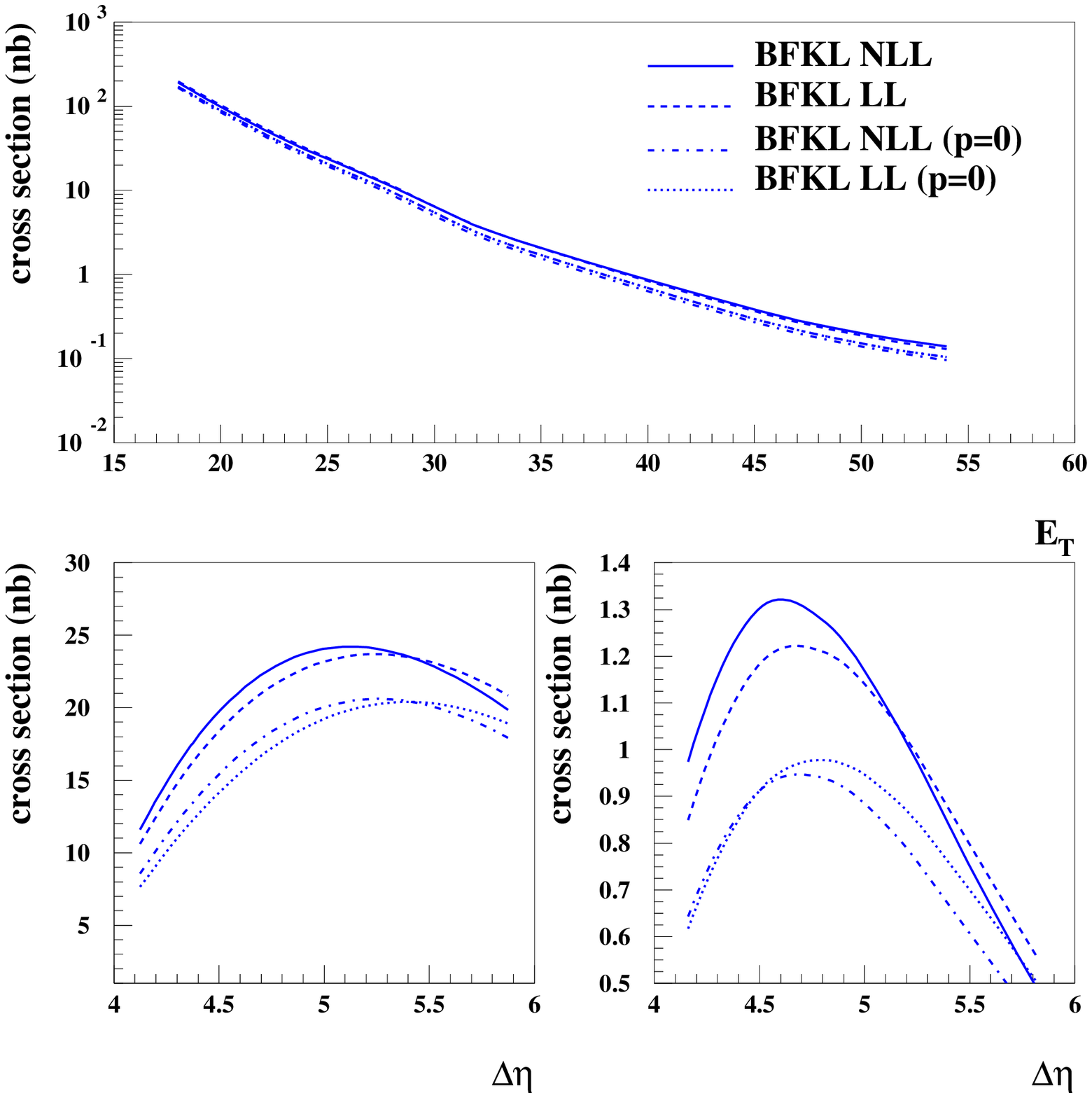,width=8.5cm}
\hfill
\epsfig{file=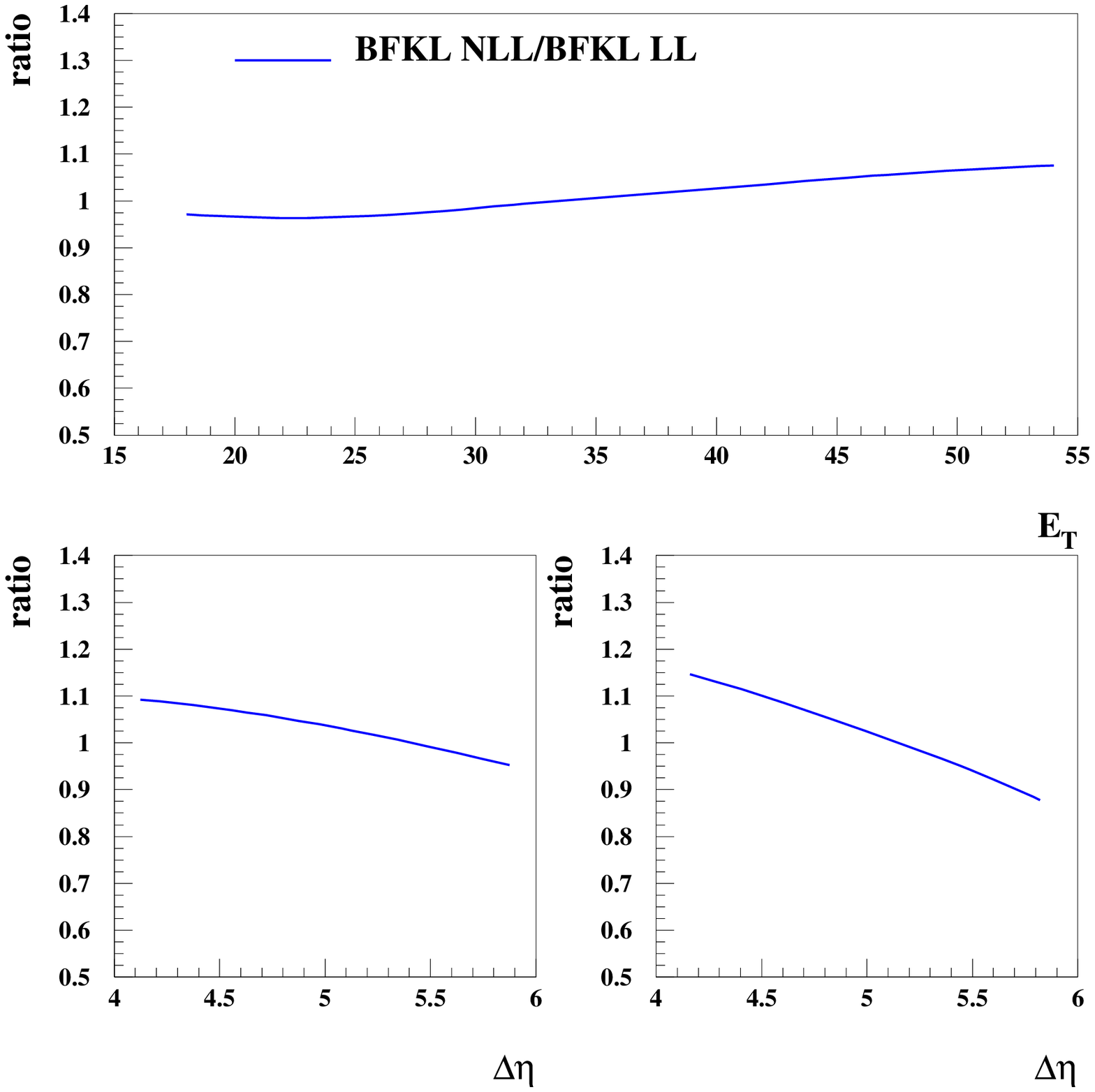,width=8.5cm}
\caption{Left plots: LL- and NLL-BFKL cross sections as a function of the momentum transfer $E_T$ (upper plot) and the rapidity gap 
$\Delta\eta$ (lower plots, $15<E_T<25\ \mbox{GeV}$ and $E_T>30 \ \mbox{GeV}$); we also give the cross sections when only the conformal 
spin component $p=0$ is included. Right plots: ratio of the NLL- and LL-BFKL calculations. Since both normalizations have
been adjusted to reproduce the data, the ratios are close to 1.}
\label{fig3}
\end{center}
\end{figure}

Let us now evaluate the relative NLL corrections with respect to the LL-BFKL cross section. In Fig.~\ref{fig3}, both calculations are 
compared with their normalizations adjusted to describe the data (this will be discussed next), therefore the cross sections displayed 
on the left plots are similar. Considering the shapes of the curves, the differences between the NLL and LL calculations are in general 
small. This is due to the small effective value used for $\bar\alpha$ in the LL calculation. However one can see on the right plots, in 
which the NLL/LL ratios are displayed, that the differences are sizeable in the case of the $\Delta\eta$ dependence for high-$E_T$ jets 
($E_T>30 \ \mbox{GeV}$).

\subsection{Comparison with the D0 measurement}

In Fig.4, we compare our predictions with the D0 measurements. The jet-gap-jet cross section is computed
using the LL- or NLL-BFKL formalism while the total cross section for dijets separated by an rapidity interval
$\Delta \eta$ is calculated using the NLOJet++ program \cite{nlojet}. To obtain the NLOQCD predictions of
the dijet cross section, by which the jet-gap-jet cross section is divided, the parton-parton hard cross
section is computed at next-to-leading order with respect to $\alpha_s,$ and the leading and next-leading logarithms 
$\alpha_s^n\log^n{E_T^2}$ and $\alpha_s^n\log^{n-1}E_T^2$ are resummed using the DGLAP equations \cite{dglap}
which govern the evolution of the parton distribution functions (PDFs) of the proton. We used the CTEQ6.1M parton distribution
functions \cite{cteq}, and the renormalization and factorization scale $\mu_r$ and $\mu_f$ were chosen as
$\mu_r=\mu_f=E_{T1},$ the transverse energy of the leading jet.

It is worth pointing out that these calculations are performed at parton level, and the hadronization effects cannot
be taken into account at this stage. However, the fact that the D0 collaboration measured ratios of cross sections
reduces the influence of hadronization corrections and of the choice for the jet algorithm, since most of these effects are the
same for jet gap jet or dijet events. This is however not the case for underlying events corrections which do not play any
role for jet-gap-jet events since a gap is observed. These corrections depend only smoothly on $E_T$ and $\Delta\eta$ in the limited
range of $E_T$ of the D0 measurement,
and their largest effect is taken into account in our approach with the gap-survival probability of ${\cal S}=0.1$ at the Tevatron.

Since the normalizations of our BFKL calculations are not under control, we fit them to the D0 measurement for the jet-gap-jet
cross section ratio as a function of $E_T.$  We take the BFKL LL or BFKL NLL formalism for the jet gap jet cross section and the NLO
QCD calculation for the inclusive cross section. 
For the NLL-BFKL (resp. LL-BFKL) calculation, the normalisation is 1.0 (resp. 0.84)
with a $\chi^2/dof$ of 0.9/6 (resp. 1.7/6) when one considers only the points with $E_T>25$ GeV in the fit. Interestingly enough, the 
normalizations are both close to 1, and our NLL-BFKL prediction with leading-order impact factors has the correct normalization and does not 
need to be adjusted after all. It is worth also noticing that the normalisations for the BFKL LL or the BFKL NLL calculations are similar.
It is quite contrary for the QCD calculation of the jet cross section when both jets are separated with a interval in rapidity
larger than $\Delta \eta$. The NLO QCD result is about 20 times larger than the LO one and it would be useful to know the QCD cross section at
NNLO to check if the higher order corrections are still large. 
The resulting jet-gap-jet cross section ratios are displayed in the upper plot of Fig.~\ref{fig4}. We notice that the 
$E_T$ dependence of the cross section is well described, with the exception of the lowest $E_T$ point where the measurement is lower than 
our calculation (this is why we did not include it in the fit to compute the normalisation). This lowest $E_T$ point is definitely more 
sensitive to hadronization or underlying event corrections, and this might be the reason why it is poorly described. We use the same 
normalisations to predict the rapidity dependence of the jet-gap-jet cross section ratio for the low- and high$-E_T$ samples and the 
result is shown on the lower plots of Fig.~\ref{fig4}. We also notice a good agreement between our calculations and the D0 measurement, 
without any further adjustment of the normalizations. This is a powerful result, the NLL-BFKL prediction has the correct normalization 
for the three measurements.

\begin{figure}
\begin{center}
\epsfig{file=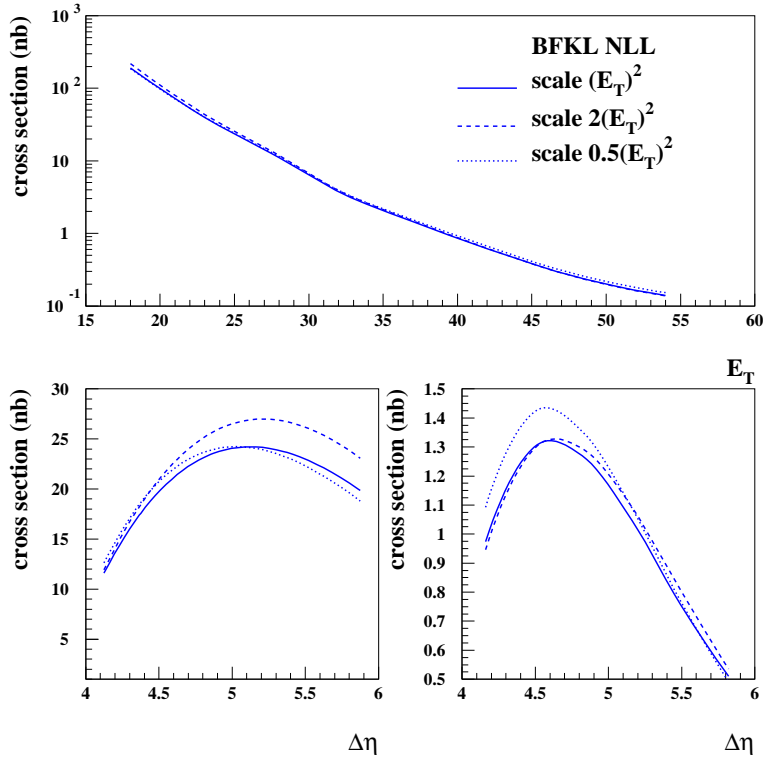,width=11cm}
\caption{Scale dependence uncertainty of the BFKL NLL calculations. The scale
uncertainty is evaluated by modifying the $E_T^2$ scale used by default to $E_T^2/2$ or $2 E_T^2$.
The effect of the scale uncertainty is of the order of 10-15\%.}
\label{fig3b}
\end{center}
\end{figure}

\begin{figure}
\begin{center}
\epsfig{file=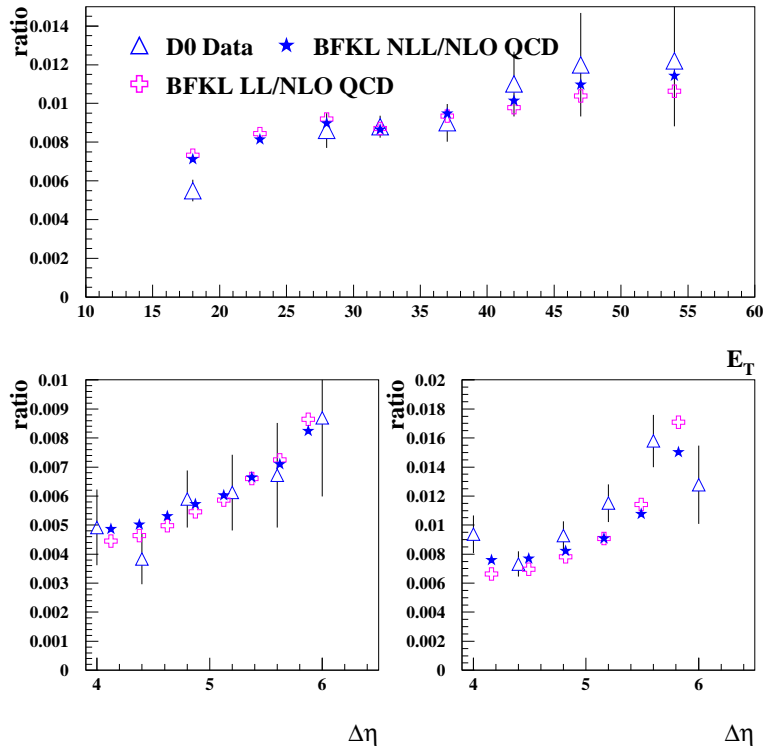,width=11cm}
\caption{Comparison between the D0 measurement of the jet-gap-jet event ratio with the NLL- and LL-BFKL calculations. The NLL calculation 
is in good agreement with the data, without the need to adjust the normalization. The LL calculation gives also a good description of the 
data after its normalization has been adjusted, although the fit shows that the NLL description is better.}
\label{fig4}
\end{center}
\end{figure}

In Fig.~\ref{fig3b}, we give the scale dependence uncertainty of the BFKL NLL calculations. As we mention in Section II, the scale
uncertainty is evaluated by modifying the $E_T^2$ scale used by default to $E_T^2/2$ or $2 E_T^2$. The uncertainty of the BFKL NLL prediction
is of the order of 10-15\%.

The fact that the LL-BFKL cross section is in good agreement with the data when a fixed value of the coupling is considered is consistent 
with the results of \cite{cfl}. The fit performed on the data as a function of $E_T$ showed that including NLL corrections in the BFKL 
framework improves the description of the data. This is also consistent with the findings of \cite{rikard}, where some NLL corrections 
were effectively included in a numerical way. In this study we confronted for the first time predictions obtained with the full analytic 
expression of the NLL-BFKL kernel to the Tevatron data, and the result is satisfactory.

\section{Predictions for the LHC}

\begin{figure}[t]
\begin{center}
\epsfig{file=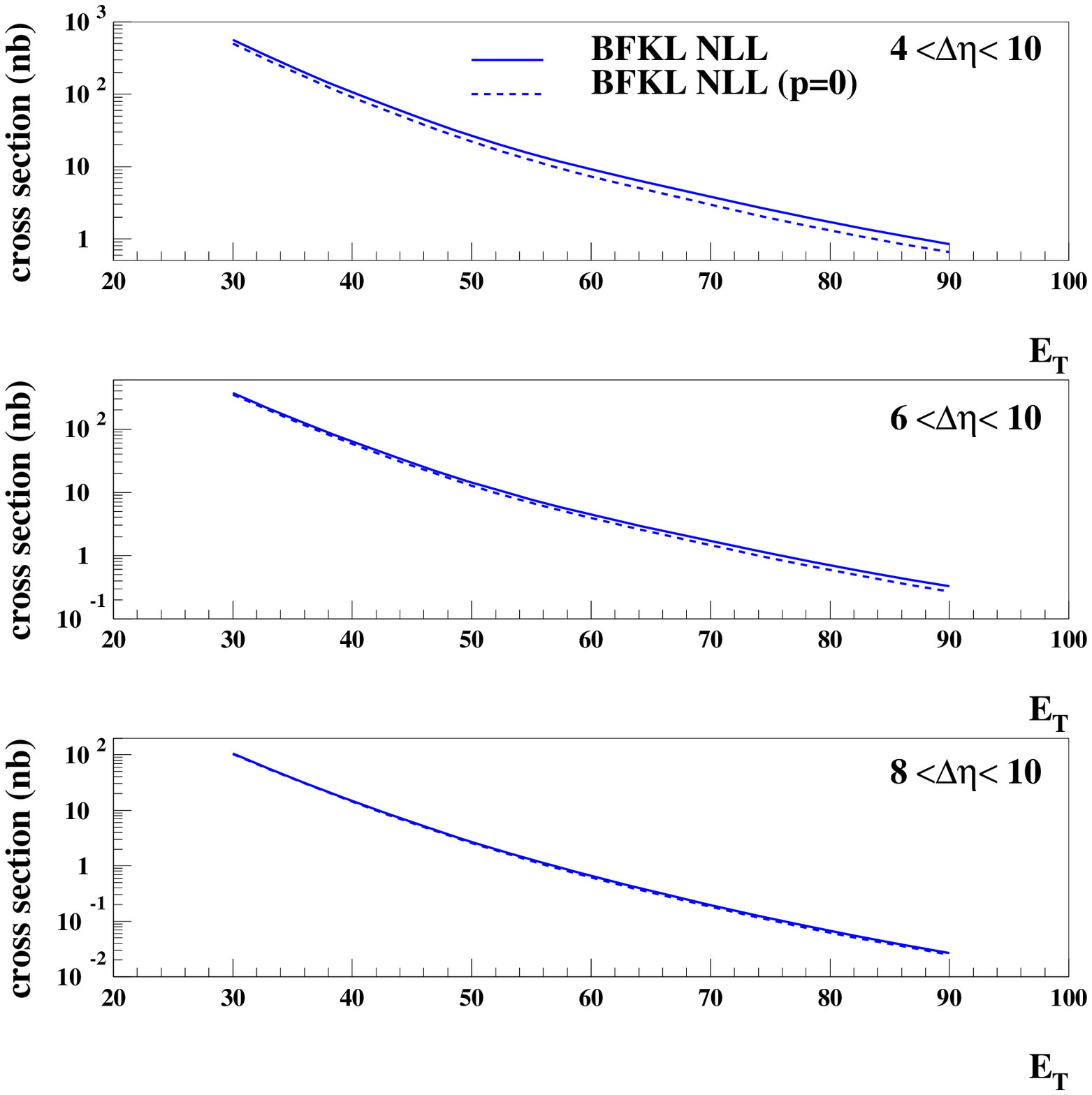,width=8.5cm}
\hfill
\epsfig{file=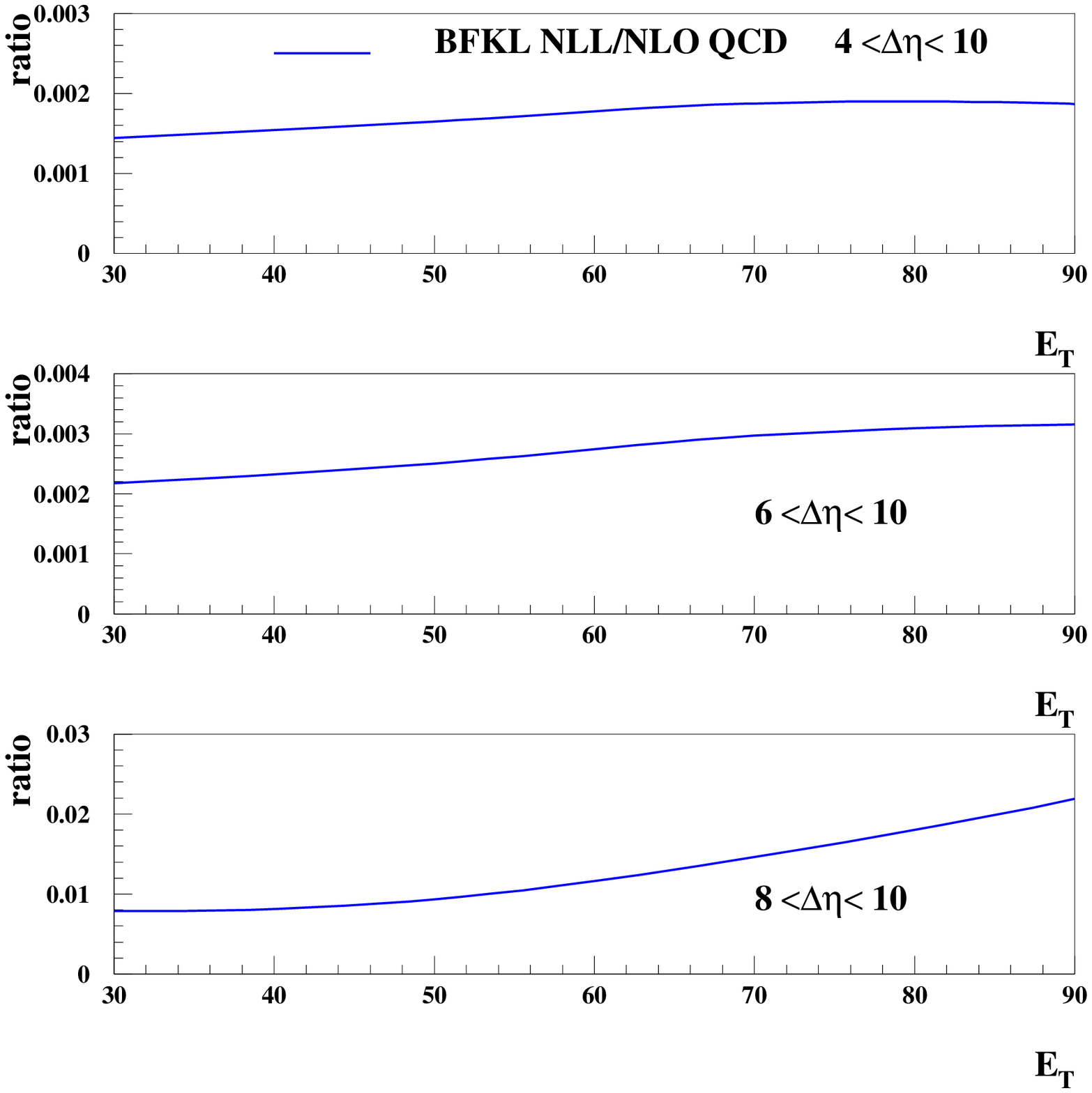,width=8.5cm}
\caption{Left plot: NLL-BFKL jet-gap-jet cross section as a function of $E_T$ for three different regions in $\Delta\eta$ at the LHC.
Right plot: the left-plot curves are divided by the NLO QCD dijet inclusive cross section.}
\label{fig5}
\end{center}
\end{figure}

\begin{figure}
\begin{center}
\epsfig{file=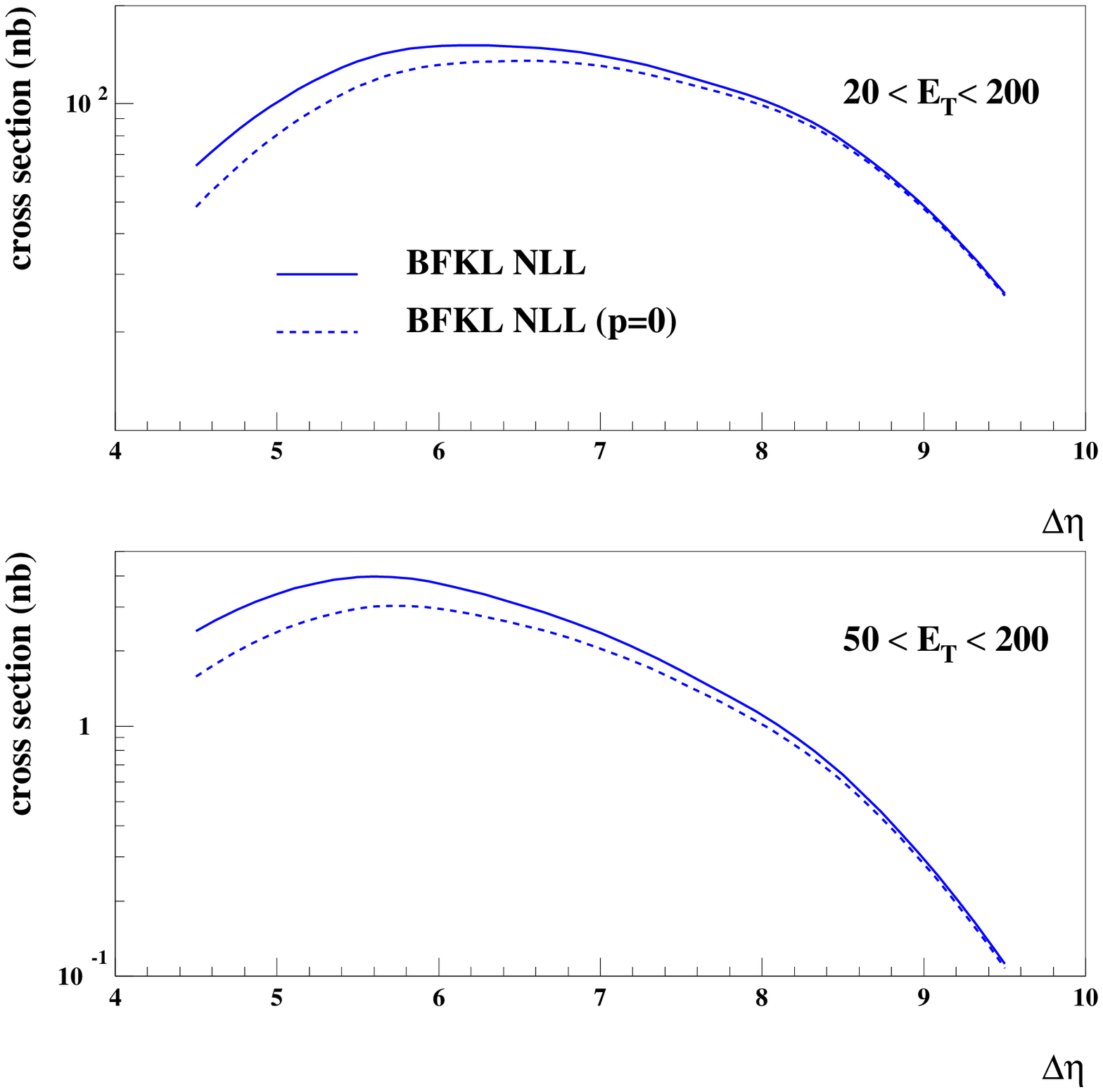,width=8.5cm}
\hfill
\epsfig{file=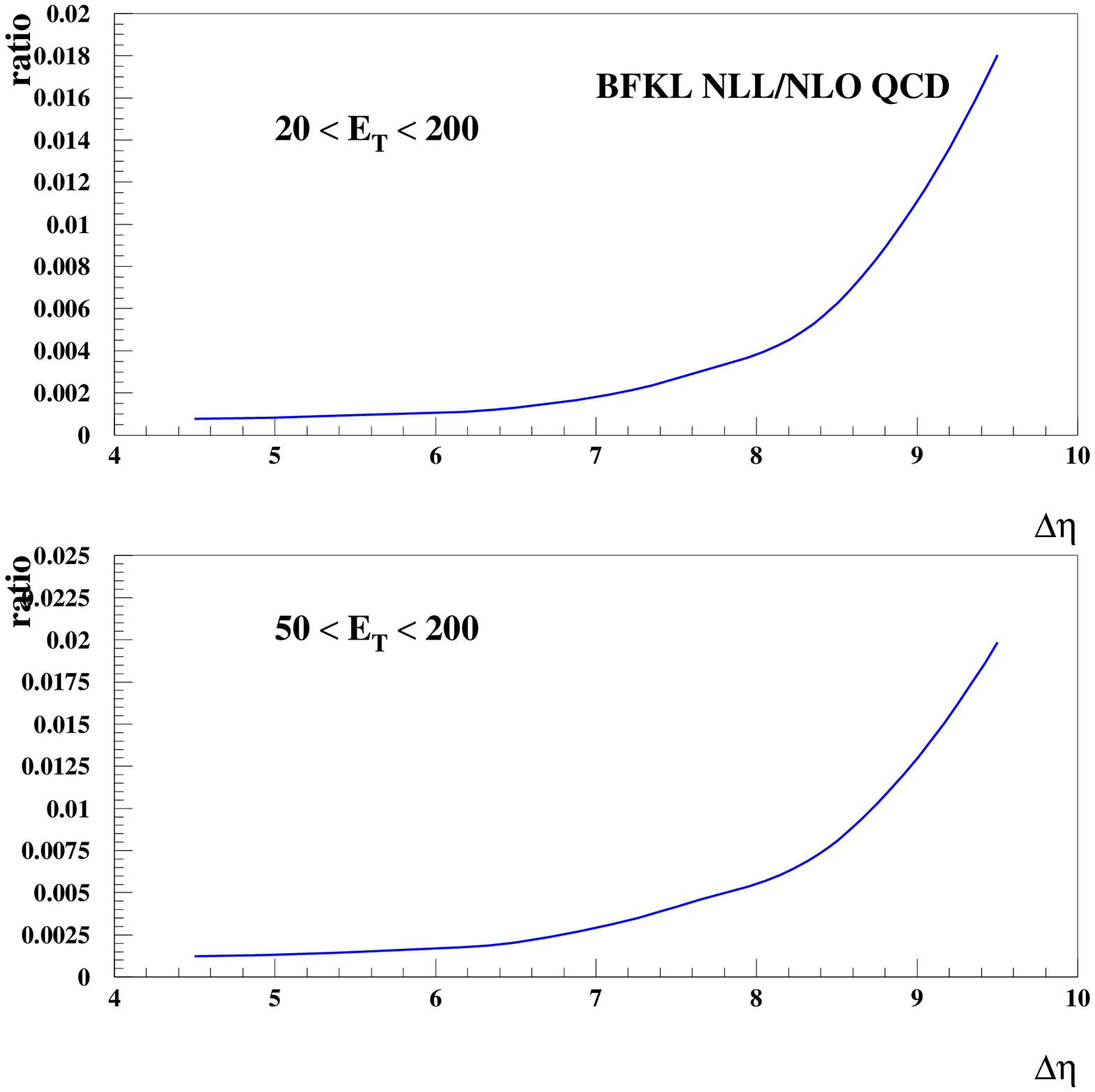,width=8.5cm}
\caption{Left plot: NLL-BFKL jet-gap-jet cross section as a function of the rapidity gap $\Delta\eta$ for two different $E_T$ regions at the 
LHC. Right plot: the left-plot curves are divided by the NLO QCD dijet inclusive cross section.}
\label{fig6}
\end{center}
\end{figure}

\begin{figure}
\begin{center}
\epsfig{file=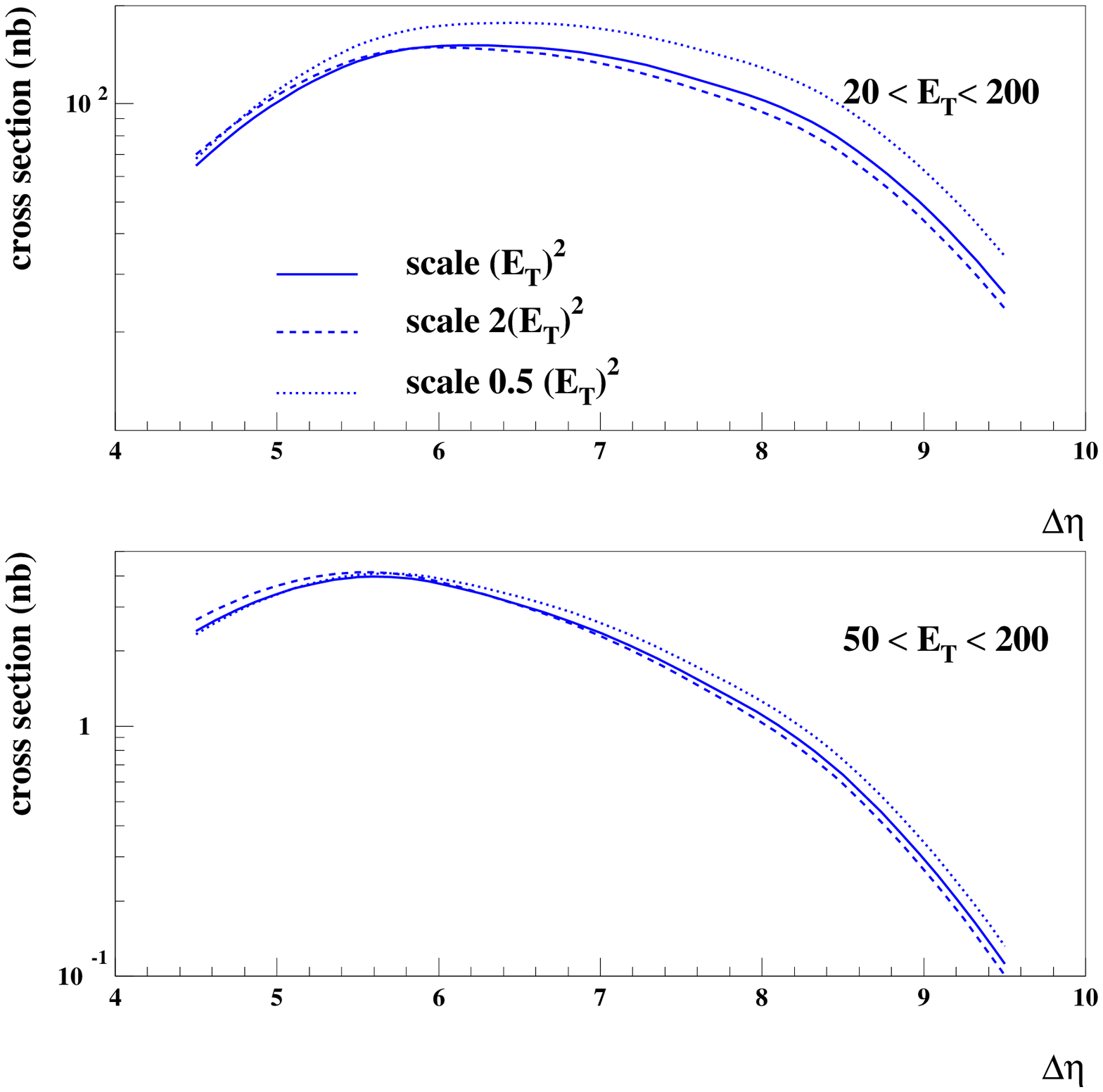,width=11cm}
\caption{Scale dependence uncertainty of the BFKL NLL calculations. The scale
uncertainty is evaluated by modifying the $E_T^2$ scale used by default to $E_T^2/2$ or $2 E_T^2$.
The effect of the scale uncertainty is of the order of 10-15\%.}
\label{fig6b}
\end{center}
\end{figure}

Using the normalizations obtained from the fits to the D0 data, we are able to predict the cross sections at the LHC.
In doing so, we also have to change the gap-survival probability from 0.1, the value used for the Tevatron, to 0.03, the proper
value for the LHC. The values of the NLL-BFKL cross sections are shown in Fig.~\ref{fig5} as a function of $E_T$ for different $\Delta\eta$
ranges, and in Fig.~\ref{fig6} as a function of $\Delta\eta$ for different $E_T$ ranges. In both figures, the cross sections are displayed
on the left plots, and on the right plots they are divided by the NLOJet++ prediction for the inclusive dijet cross section.
In Fig.~\ref{fig6b}, we display the scale uncertainties of the BFKL NLL calculation by modifying the default $E_T^2$ scale
to $E_T^2/2$ or $2 E_T^2$ as a function of the $\Delta \eta$ between the two jets (the uncertainty is the same
as a function of jet $E_T$). The uncertainty of the BFKL NLL prediction is of the order of 10-20\%.

\begin{figure}[t]
\begin{center}
\epsfig{file=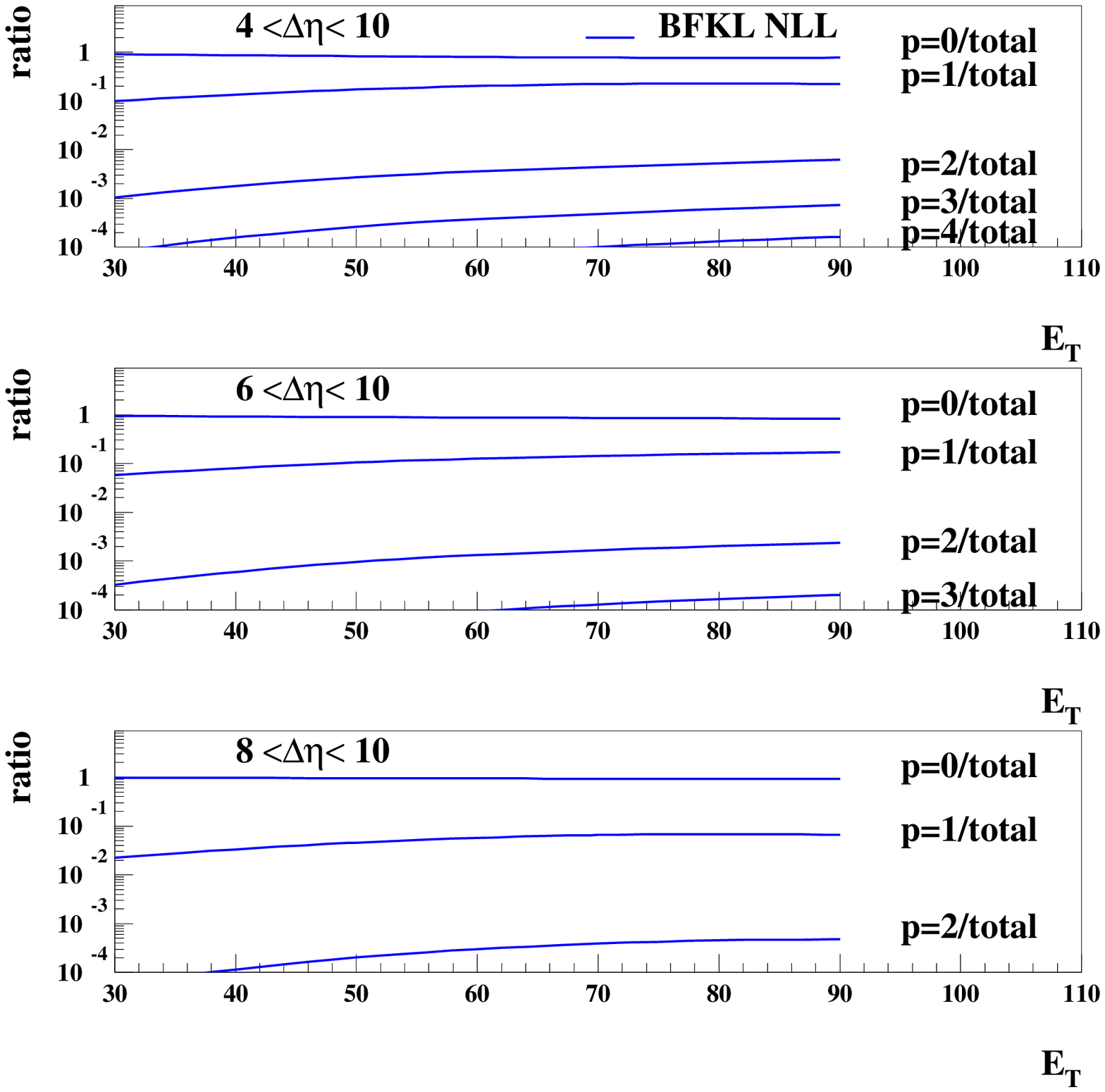,width=8.5cm}
\hfill
\epsfig{file=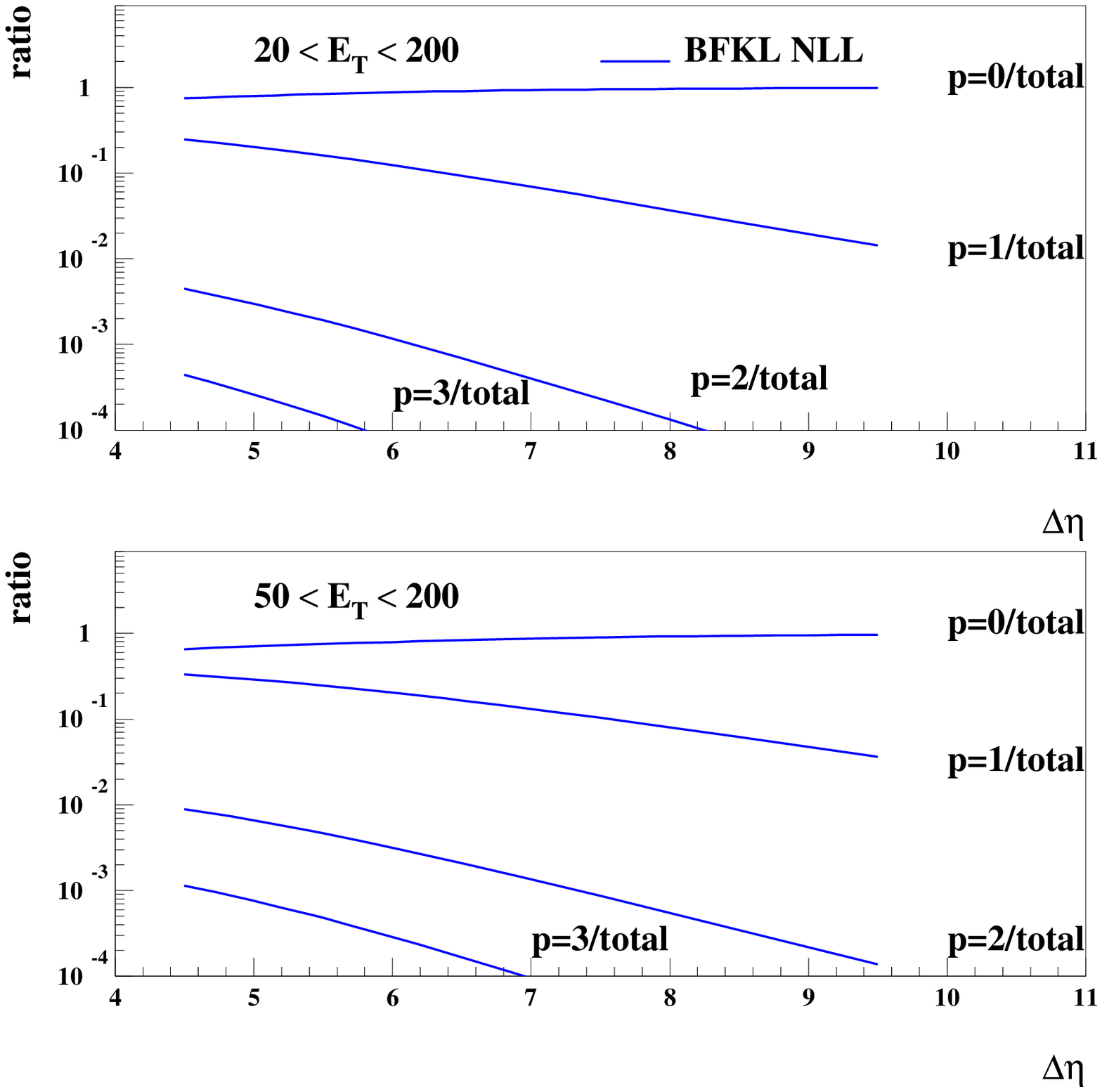,width=8.5cm}
\caption{Ratio between the conformal-spin components $p=0$, 1, 2, 3 and 4, and the total BFKL cross section at NLL accuracy. Left plot: as 
a function of $E_T$ in different regions of $\Delta \eta;$ right plot: as a function $\Delta\eta$ for a leading jet with 
$20<E_T\ <200\mbox{GeV}$ and $50<E_T\ <200\mbox{GeV}.$}
\label{fig7}
\end{center}
\end{figure}

Comparing the relative contributions from the different conformal spin components, we could see that the $p=1$ and
$p=2$ components are large at small $\Delta\eta$ and at large $E_T,$ and they cannot be neglected as it is shown in
Fig.~\ref{fig7}. At large $\Delta\eta$
and low $E_T$, the $p=0$ component is by far the largest one. Also, it can be noticed that the cross section is high enough
to be measured at the LHC in special runs, collecting integrated luminosities of a few 100 pb$^{-1}.$ Nevertheless, a good energy calibration will be needed to deal with not so energetic jets which are in addition quite forward. Finally, a clear prediction is that
the percentage of the jet-gap-jet events is higher when the interval in rapidity between the two jets $\Delta \eta$ is large.

\section{Conclusions}

Within the BFKL framework at NLL accuracy, we have investigated diffractive events in hadron-hadron collisions in which two high$-E_T$ jets 
are produced and separated by a large rapidity gap $\Delta\eta.$ Using renormalization-group improved NLL kernels in the S4 scheme, the 
NLL-BFKL effects were taken into account through the effective kernel obtained from the implicit equation \eqref{eff}. We implemented the 
MT prescription to couple the BFKL Pomeron to colored impact factors, this allowed our phenomenological study of NLL-BFKL effects in 
jet-gap-jet events. We point out that only used leading-order impact factors, the implementation of the NLO impact factors goes beyond 
the scope of our phenomenological analysis.

Dividing our jet-gap-jet BFKL calculations by the NLO QCD preditions for the inclusive dijet cross section, we showed that the BFKL 
predictions were in good agreement with the Tevatron data for both the $E_T$ and $\Delta\eta$ dependence of the jet-gap-jet cross section 
ratios. In the case of the NLL calculation, adjusting the normalization was not needed for any of the three measurements. The LL-BFKL 
predictions can also describe the data with the fixed value of the coupling $\bar\alpha=0.16,$ and a normalization factor of order one. Still the better description was obtained with the NLL formulation. We presented predictions which could be tested at the LHC, for the same jet-gap-jet event ratio measured at the Tevatron, but for larger rapidity gaps which can be obtained at the LHC. This should provide a strong test of the BFKL regime.

\begin{acknowledgments}

CM is supported by the European Commission under the FP6 program, contract No. MOIF-CT-2006-039860.

\end{acknowledgments}

\end{document}